\documentclass[prd, 11pt, a4paper, showpacs, notitlepage, dvipsnames, reprint, twocolumn, preprintnumbers, nofootinbib]{revtex4-1}

\usepackage{geometry}
\usepackage[utf8x]{inputenc}
\usepackage{makecell}
\usepackage{float}
\usepackage{tabularx}
\usepackage[activate={true,nocompatibility},final,tracking=true,kerning=true,factor=1500,stretch=10,shrink=10]{microtype}
\usepackage{graphicx}
\usepackage{bm, amssymb, amsmath, amsfonts}
\usepackage{multirow}
\usepackage{xcolor}

\usepackage{url}

\usepackage{url}
\usepackage[breaklinks]{hyperref}

\usepackage{hyperref}
\hypersetup{
colorlinks = true, 	    
linkcolor = magenta,	
citecolor = blue,		
}

\usepackage[linesnumbered,ruled,lined,boxed]{algorithm2e}
\usepackage[noend]{algpseudocode}

\SetAlCapSty{xAlCapSty}

\SetCommentSty{mycommfont}

\SetNlSty{mynlfont}{}{} 

\RestyleAlgo{algoruled}

\usepackage{subcaption}

\setcitestyle{numbers,square}

\begin{document}

\preprint{LIGO DCC number \bf LIGO-P2100087}

\title{
Convolutional neural networks for the detection of the early \break
inspiral of a gravitational-wave signal
}

\author{Grégory Baltus$^{1}$}	\email[]{gbaltus@uliege.be}
\author{Justin Janquart$^{2,3}$}	    \email[]{j.janquart@uu.nl}
\author{Melissa Lopez$^{2,3}$}	\email[]{m.lopez@uu.nl}
\author{Amit Reza$^{2,3}$} \email[]{areza@nikhef.nl}
\author{Sarah Caudill$^{2,3}$} \email[]{s.e.caudill@uu.nl}
\author{Jean-René Cudell$^{1}$} \email[]{jr.cudell@uliege.be}

\affiliation{${}^1$ STAR Institut, Bâtiment B5, Université de Liège, Sart Tilman B4000 Liège, Belgium}
\affiliation{${}^2$ Nikhef, Science Park 105, 1098 XG Amsterdam, The Netherlands}
\affiliation{${}^3$ Institute for Gravitational and Subatomic Physics (GRASP), Utrecht University, Princetonplein 1, 3584 CC Utrecht, The Netherlands}

\begin{abstract}
GW170817 has led to the first example of multi-messenger astronomy with observations from gravitational wave interferometers and electromagnetic telescopes combined to characterise the source. However, detections of the early inspiral phase by the gravitational wave detectors would allow the observation of the earlier stages of the merger in the electromagnetic band, improving multi-messenger astronomy and giving access to new information. In this paper, we introduce a new machine-learning-based approach to produce early-warning alerts for an inspiraling binary neutron star system, based only on the early inspiral part of the signal. We give a proof of concept to show the possibility to use a combination of small convolutional neural networks trained on the whitened detector strain in the time domain to detect and classify early inspirals. Each of those is targeting a specific range of chirp masses dividing the binary neutron star category into three sub-classes: light, intermediate and heavy. 
In this work, we focus on one LIGO detector at design sensitivity and generate noise from the design power spectral density. We show that within this setup it is possible to produce an early alert up to 100 seconds before the merger for the best-case scenario. 
We also present some future upgrades that will enhance the detection capabilities of our convolutional neural networks. Finally, we also show that the current number of detections for a realistic binary neutron star population is comparable to that of matched filtering and that there is a high probability to detect GW170817- and GW190425-like events at design sensitivity. 
\end{abstract}
\maketitle 
\section{Introduction}
\label{Sec:Intro}

On August 17, 2017 the first gravitational wave (GW) from a binary neutron star 
(BNS) system was observed by the Laser Interferometer Gravitational Wave 
Observatory (LIGO)~\cite{TheLIGOScientific:2014jea} and by the Virgo 
detector~\cite{TheVirgo:2014hva}. 

The Fermi Gamma-Ray-Burst Monitor (Fermi-GBM) \cite{meegan2009fermi} and the INTEGRAL satellite \cite{savchenko2017integral} detected the associated $\gamma$-ray  signal 1.7 s after the coalescence. This event, called GW170817, provided the first direct evidence of a link between these mergers and short $\gamma$-ray bursts. In addition, it gave an extra confirmation of the existence of GWs and initiated the era of multi-messenger astronomy (MMA) with GWs~\cite{TheLIGOScientific:2017qsa, abbott2019low, abbott2017multi}. 
The combined detection of multiple messengers allows us to improve our understanding of complex astrophysical phenomena, such as the r- and s-processes at the origin of heavier elements in the Universe. This improvement of MMA will also allow a better measurement of the Hubble constant and novel tests of General Relativity, such as a measurement of the speed of GWs \cite{cowperthwaite2017electromagnetic, soares2019first, fishbach2019standard, berti2018extreme, abbott2019tests, abbott551hubble}. A key element in MMA is the time delay between the detection of a GW and the identification of the location of its source. In fact, to discover new physics it is necessary to detect the source in the electromagnetic band at times close to the merger. However, the time of response varies from one telescope to another, so that a detection in the GW channels should be made early to leave enough time for the electromagnetic observatories to focus on the source. For instance, the Swift Observatory~\cite{Burrows:2005gfa} is able to focus on a sky position in only seconds (around 15 s for Swift's Burst Alert Telescope~\cite{Burrows:2005gfa}). Thus, it is advantageous to detect gravitational waves from the inspiral, before the merger, to enable prompt detection of the merger event in the electromagnetic band.


The signal of a GW coming from compact binary coalescence (CBC) is composed of three parts: the inspiral (when the orbital motion of the two objects radiates away energy and the orbit shrinks), the merger (when they touch and join), then by a ringdown (when the newly formed body returns to its ground state). As the signal enters in the detector, the signal-to-noise-ratio (SNR) accumulates. Due to the low frequency and small amplitude of the signal during the early inspiral,  the SNR accumulation is slow, which hinders the detection of the signal. It becomes observable when the frequency of the early inspiral enters into the sensitivity band of the detector. Upgrading the current  detectors and building the next generation of interferometers, such as Cosmic Explorer (CE) and the Einstein Telescope (ET)~\cite{Reitze:2019iox,Punturo:2010zz,Sathyaprakash:2012jk},  will increase significantly the sensitivity through, among other things, the reduction of the noise. As a consequence the frequency threshold will also decrease, leading to the detection of signals with longer inspirals and to a larger number of detections.  Up to now, both the low-amplitude early inspiral and the high-amplitude late inspiral have been needed to detect BNSs.


The standard methodology employed to search for GWs relies on 
\textit{matched filtering} techniques. A large bank of template waveforms is 
built. The templates are then correlated with the input data of the detector over 
their sensitivity band, extracting the signals from the detector noise. The 
standard matched filtering can be computationally expensive 
~\cite{cannon2012toward,usman2016pycbc, dal2014implementing, 2017PhRvD..95d2001M},
but some pipelines such as GstLAL~\cite{sachdev2019gstlal}, PyCBCLive 
~\cite{nitz2018rapid}, MBTAOnline ~\cite{adams2016low} and SPIIR 
~\cite{Chu:2020pjv} are adapted to run in low latency and obtain fast candidate detection, also referred to as trigger.


Matched filtering techniques usually consider the whole template (meaning the template over all the sensitive frequencies of the detector) to correlate it with the signal. However in the context of MMA it is necessary to employ only the pre-merger information of the template for prompt alerts. In this line of thought, recent advances to perform matched filtering with only a fraction of the inspiral have been made in Ref.~\cite{Sachdev:2020lfd}, where the authors have implemented a GstLAL-based pipeline that produces pre-merger alerts. 

It enabled them to compute the matched filter, the false-alarm rate, and the sky localisation using only the information in the low-frequency band of the template, corresponding to the early inspiral. This showed that this method could detect signals as early as one minute before the merger. However, the number of early alerts issued is lower than the total number of detections based on full waveforms.

Due to the computational complexity of matched filtering and the increasing amount of events related to the future upgrades of the detectors, alternative approaches to overcome the challenges of MMA are under development. In particular, the use of Machine Learning (ML) methods has sparked the interest of several authors, who have built Deep Learning (DL) algorithms. These algorithms are able to capture complex non-linear relationships in the data by composing hierarchical internal representations. The main advantage of these methodologies is that the prediction task is performed rapidly since most of the computations are made during the training stage ~\cite{George:2016hay}. Several studies have shown the power of these algorithms for the detection of GW in low latency, obtaining a sensitivity similar to that of matched filtering techniques ~\cite{George:2016hay, George:2017pmj, Gabbard:2017lja, wei2020deep, krastev2020real, gebhard2017convwave}. Other recent papers \cite{gabbard2019bayesian, green2020gravitational, green2020complete, cuoco2020enhancing} focus on the parameter estimation for CBC events and other applications of ML for gravitational wave astronomy. Moreover, in ~\cite{George:2016hay} and ~\cite{George:2017pmj} the authors have presented the generalisation ability of convolutional neural networks (CNNs) by training with a data set of non-spinning waveforms and obtaining a high performance when testing with precessing systems.

In~\cite{wei2020deep}, only part of the template is used in their DL approach. The authors employed a pre-trained  Resnet-50 network~\cite{he2016deep} to classify time-frequency maps. The data were acquired from the detectors after an extra preprossessing step to build a spectrogram of the data. When computing such figures, depending on the desired resolution, we found that the time to produce a single map could vary from $\sim$ 0.5 seconds \footnote{All the tests on CPU were done on a Intel(R) Core(TM) i7-8650U CPU.}.

In this paper, we propose to use the 1-D whitened strain as the input data of a CNN for pre-merger alert so that we bypass the computation of the spectrograms. The goal of the algorithm is to perform a binary classification task to differentiate inputs that contain a GW from inputs that do not.  The classification is made independently for three different categories of objects: light, intermediate and heavy BNS. The templates with a GW contain only the early inspiral part of the waveform, which is embedded in colored Gaussian noise, made with the noise power spectral density (PSD), corresponding to the design sensistivity of LIGO and given in PyCBC~\cite{Biwer:2018osg}. Different categories have different observational time windows (OTW), i.e. the duration of inspiral seen by each CNN is different depending on the category. A short CNN, as in ~\cite{George:2016hay, George:2017pmj, Gabbard:2017lja}, is implemented for each category. We stress that this work is a proof of concept to show the promises of this type of neural networks. The optimisation of their performance and the inclusion of multiple detectors will be considered in further studies.

This paper is organised as follows. In the first section, we discuss the data generation. The second section details the methodology used to design and train the networks. The third section is devoted to our different results and discusses them. Finally, we draw our conclusions and consider future enhancements to be brought to our detection system. 


\section{Data Generation}
\label{sec:PartaiSNR}
\subsection{SNR and Partial-Inspiral SNR}

The final output of a matched filtering algorithm is the signal-to-noise ratio (SNR). It measures the match between the template and the data. Mathematically, the SNR ($\rho$)~\cite{PhysRevD.85.122006} is defined as 
\begin{equation}\label{eq:SNR}
    \rho = \left(4\, \mathfrak{Re}\left(\int_{f_{min}}^{f_{max}}\frac{{\tilde d}(f)\tilde{h}^{*}(f)}{\mathcal{P}(f)}df\right)\right)^{1/2} \, ,
\end{equation}
where $\tilde{h}^{*}(f)$ and $\tilde{d}(f)$ are respectively the 
complex conjugate of the Fourier transform of the template and the Fourier 
transform of the data. $\mathcal{P}(f)$ is the noise power spectral density. Here, $f_{min}$ is the minimal frequency in the detector sensitivity band and $f_{max}$ is the maximum frequency considered, typically the Nyquist frequency, i.e. half of the sampling frequency.

The SNR represents how well a typical template $h$ matches the data $d$, which is the addition of a GW signal and noise. A matched-filtering-based search finds the template that maximises the SNR and is optimal for Gaussian stationary noise and an exactly known signal. For this type of noise, when the signal does not contain a GW, the SNR fluctuates around a mean value. Nonetheless, if a GW enters the detector, the SNR increases and when that exceeds a predefined SNR threshold, a candidate trigger event is recorded. 

However, the noise from the detectors is neither Gaussian nor stationary, making the search more complex. For example, glitches (spurious noise variations in the detector band) can occur and lead to a peak in SNR which can mimic a GW trigger. To avoid the detection of noise, the matched filtering-based pipelines often require the detection to be in coincidence in different detectors.  Additionally, more elaborate tests also exist, such as the  $\chi^{2}$ test~\cite{Allen:2004gu}, that can downrank the noise artefacts in the final candidate lists. The confidence one has about the detection is also often translated by a false-alarm-rate that gives the frequency at which noise fluctuations lead to the same ranking statistic value. This ranking is a multi-variate statistic that includes different statistics such as SNR and $\chi^{2}$~\cite{2017PhRvD..95d2001M}.


The optimal SNR is obtained when the template is matched with itself~\cite{PhysRevD.85.122006}:

\begin{equation}\label{eq:optimalSNR}
    \rho_{opt} = \left(4\, \mathfrak{Re}\left(\int_{f_{min}}^{f_{max}}\frac{|\tilde{h}(f)|^2}{\mathcal{P}(f)}df\right)\right)^{1/2} \, ..
\end{equation}

This  value represents the loudness of the signal in the detector. 

In the context of pre-merger analysis, only part of the inspiral is considered, and the loudness of the signal is not represented anymore by the optimal SNR. Instead, we define it by a partial inspiral SNR (PI SNR), which has the same definition as the optimal SNR in Eq.~(\ref{eq:optimalSNR}), but where the template is now the partial template containing only the early partial inspiral ($h_{PI}$). In the frequency domain, it is equivalent, for a given waveform, to replace the $f_{max}$ in Eq.(\ref{eq:optimalSNR}) by the maximum frequency reached by the template in the part of the inspiral considered. Typically, this frequency will be below 50~Hz (instead of thousands usually), reducing the value of the integral.

\begin{figure}[ht]
\begin{center}
    \includegraphics[scale=0.7]{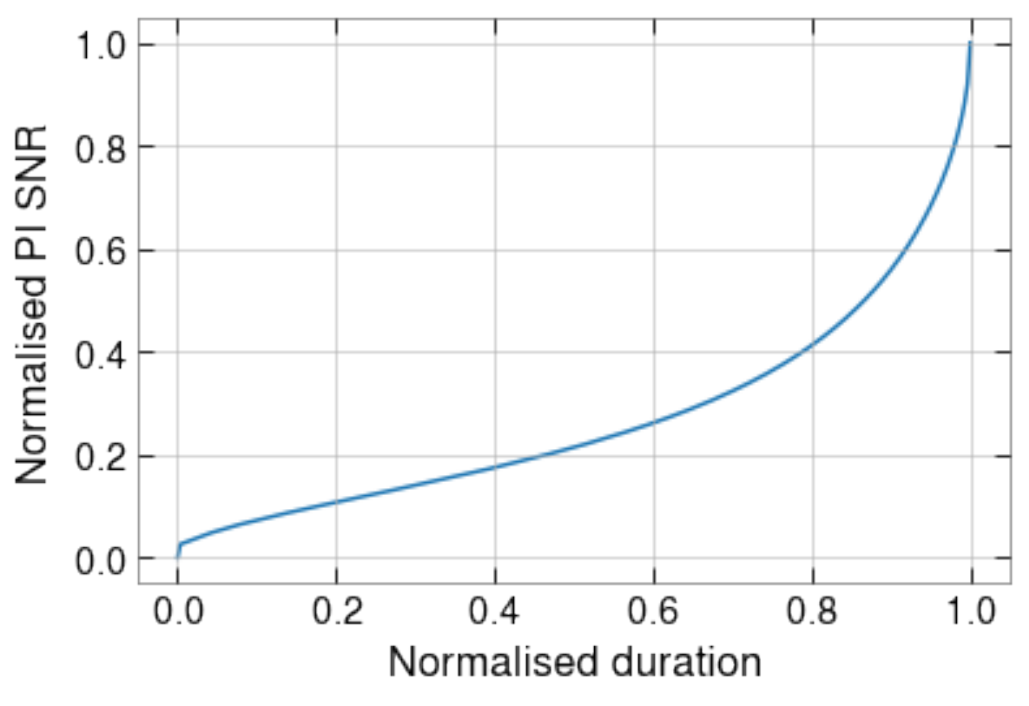}
    \caption{\label{fig:PISNR_evo} Evolution of the PI SNR as a function of the duration of the early inspiral for a BNS with component masses of 1~$M_{\odot}$. On the vertical axis, the PI SNR is normalised by the optimal SNR, and on the horizontal axes, duration of the early inspiral is normalised by the duration of the full template.} 
\end{center}
\end{figure}

The SNR increases more rapidly around the late inspiral and the merger than during the early inspiral. Fig.~\ref{fig:PISNR_evo} shows this behaviour as it represents the value of the PI SNR as a function of the fraction of the signal that is taken into account.

\begin{figure}[ht]
\begin{center}
    \includegraphics[keepaspectratio, width=0.45\textwidth]{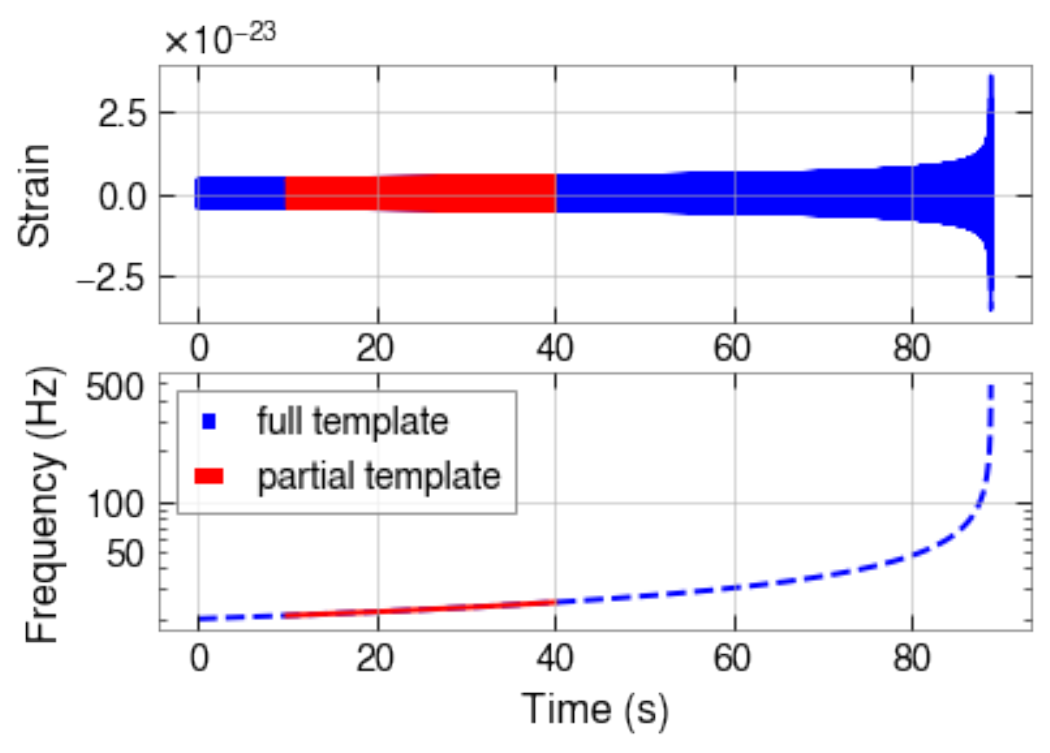}
    \caption{\label{fig:partial_vs_full} The top panel represents an intermediate BNS template where the component masses are $m_{1} = m_{2} = 2 M_{\odot}$. The bottom panel shows the frequency evolution of the template with time. In both plots, the inspiral part considered for our ML-based approach is coloured in red.}
\end{center}
\end{figure}

The behaviour of the PI SNR comes from the relation between frequency and time. At the lowest order in velocity, one finds:
\begin{equation}
    f(t) = \frac{1}{\pi} \bigg(\frac{G \mathcal{M}_{c}}{c^{3}}\bigg)^{-5/8} \bigg(\frac{5}{251}\frac{1}{(t_{m}-t)}\bigg)^{3/8},
\end{equation}
where $f(t)$ is the frequency at time $t$, $\mathcal{M}_{c}$ is the chirp mass defined in terms of the component masses $m_{1}$ and $m_{2}$ of the system:  $\mathcal{M}_{c}=(m_1 m_2)^{3/5}/(m_1+m_2)^{1/5}$ and $t_{m}$ is the time of the merger. This behaviour is illustrated in Fig.~\ref{fig:partial_vs_full}, which shows the full and partial templates and their frequency evolution.

\subsection{BNS categories}

The duration of the observable CBC signal depends mainly on the chirp mass $\mathcal{M}_{c}$. 
Indeed, at the lowest order in velocity\footnote{In the early inspiral, the strong field effects and the velocities are rather small, which means that the expression derived for the lowest order in velocity approximates well the behaviour of the binary systems.}, the duration of the signal is given by~\cite{Samajdar:2021egv}
\begin{equation}\label{eq:duration}
\tau(s)\simeq \frac{3}{(\mathcal{M}_{c}/M_{\odot})^{\frac{5}{3}}} \bigg[ \bigg(\frac{100~{\rm Hz}}{f_{low}}\bigg)^{\frac{8}{3}}-\bigg(\frac{100~{\rm Hz}}{f_{high}}\bigg)^{\frac{8}{3}}\bigg],
\end{equation}
where $f_{low}$ is the lowest frequency in the detector sensitivity band and $f_{high}$ is the highest frequency reached by the binary (approximated by the frequency of the innermost stable orbit). From this expression it is clear that, for a fixed lowest frequency $f_{low}$, if the chirp mass $\mathcal{M}_{c}$ increases, the duration of the detectable signal shortens. 

Furthermore, at the lowest order in velocity, the SNR also has a simple expression~\cite{Cutler:1994ys}:
\begin{equation}\label{eq:rho}
{\rm \rho} \simeq \frac{1}{2}\sqrt{\frac{5}{6}}\frac{1}{\pi^{\frac{2}{3}}}\frac{c}{D}\bigg(\frac{G\mathcal{M}_{c}}{c^{3}} \bigg)^{\frac{5}{6}}\sqrt{I}\, g(\theta, \phi, \psi, \iota).
\end{equation}

In this expression, $c$ is the speed of light, $D$ is the luminosity distance, $G$ is the gravitational constant, $\mathcal{M}_{c}$ is the chirp mass, $I$ is the frequency integral 
\begin{equation}\label{eq:FreqInt}
    I=\int_{f_{min}}^{f_{max}} \frac{(f^\prime)^{-7/3}}{P(f')} \, d f^\prime\,,
\end{equation}
 and $g(\theta, \phi, \psi, \iota)$ is a function that depends on the orientation of the orbital plane and on the sky position through the antenna pattern of the detectors~\cite{Schutz:2011tw}. From Eqs. (\ref{eq:duration}) and (\ref{eq:rho}) one can see that if the chirp mass decreases, the optimal SNR of the signal decreases while its duration increases.


As we can observe in Fig.~\ref{fig:PISNR_evo}, the PI SNR depends on the fraction of the signal considered, as well as on the highest frequency reached within the observation time. Therefore, observing the signal for a longer time would lead to a higher PI SNR, making the signal easier to detect. However, we also want to detect the signal as early as possible in order to have an efficient pre-merger alert system. This leads to a trade-off in our method, as we want to have a high PI SNR, but also prompt detections. 

Since we know that the time evolution of the amplitude of the signals will be different depending on the masses, we split the BNS set into three different categories: light, intermediate and heavy BNS. For each of these categories, we use a different observation time window (OTW), meaning that we train the networks on a different length of data. 
Hence, our algorithm consists of 3 CNNs, one for each category and input size. Note that the OTW is a hyperparameter that will be tuned in a later work. A discussion of the influence of this parameter will be discussed in section IV.


Table~\ref{tab:inputs} summarises the characteristics of the different categories, which are classified according to their chirp mass $\mathcal{M}_{c}$. In order to give an intuition for the masses of the objects present in each category, we give the highest and lowest chirp masses of each category and the component masses for an equal-mass system\footnote{Non-equal mass systems were also considered during the training and testing of our networks.}. For each of the categories, in addition to the constraint on the chirp mass, we also restricted the individual component masses to be between $1\,M_{\odot}$ and $3\,M_{\odot}$, which corresponds to a broad mass range for neutron stars. Note also that spin effects are absent at this order in $\frac{v}{c}$, so that we considered only non-spinning BNS.

\begin{table}
\begin{center}
\begin{tabular}{|c|c|c|c|}
  \hline
  BNS & light & intermediate & heavy  \\
  \hline
  \hline
  \makecell{$\mathcal{M}_{c}$ (\(M_\odot\)) }  & 1.13 - 1.56 & 1.56 - 2.09 & 2.09 - 2.61 \\\hline
  $f_{low}$ (Hz) & 20 & 20 & 20 \\\hline
  Duration (s) & 100 - 180 & 65 - 100 & 45 - 65  \\\hline
  \makecell{OTW  (s)} & 80 & 50 & 30 \\\hline
  Fraction of signal & 0.44 - 0.8 & 0.5 - 0.77 & 0.46 - 0.66\\\hline
  \makecell{Early alert \\ before merger (s)}  & 20 - 100 & 15 - 50  & 15 - 35   \\

  \hline
\end{tabular}
\end{center}
\caption{Summary of the CBC merger types for the different CNNs. A different OTW is considered for each category because of the difference in duration of the signals depending on the component masses. The fraction of signal corresponds to the time passed by the signal in the frame w.r.t the total duration. The early alert time is the  duration between the end of the OTW and the end of the signal. For each OTW, the minimum and maximum component masses are 1 and 3 $M_\odot$.}
\label{tab:inputs}
\end{table}

\subsection{Data set generation}

The inputs of the neural networks are 1-dimensional whitened time series, made of Gaussian noise generated from the design sensitivity PSD of Advanced LIGO (aLIGO) with a GW added in some cases. Indeed, the network is trained as a classifier between an \textit{event class} (noise + template) and a \textit{noise class} (only noise). The GW data analysis and generation has been performed with the PyCBC package~\cite{Biwer:2018osg}.

We start by generating 120 seconds of coloured Gaussian noise. Then, a non-spinning BNS waveform is injected into it. The approximant used is \textit{SpinTaylorT4}~\cite{Buonanno:2002fy} and it is generated with a minimum frequency of 20 Hz. 

By default, we employ the optimal sky localisation considering only the plus polarisation of the GW aligned with the arms of the interferometer to generate training, validation and testing sets. Note however that when we later will test the performance of the networks with realistic BNS populations, similarly to~\cite{Samajdar:2021egv}, the sky location will not be the optimal one anymore. 

\begin{figure}[ht]
\begin{center}
    \includegraphics[scale=0.75]{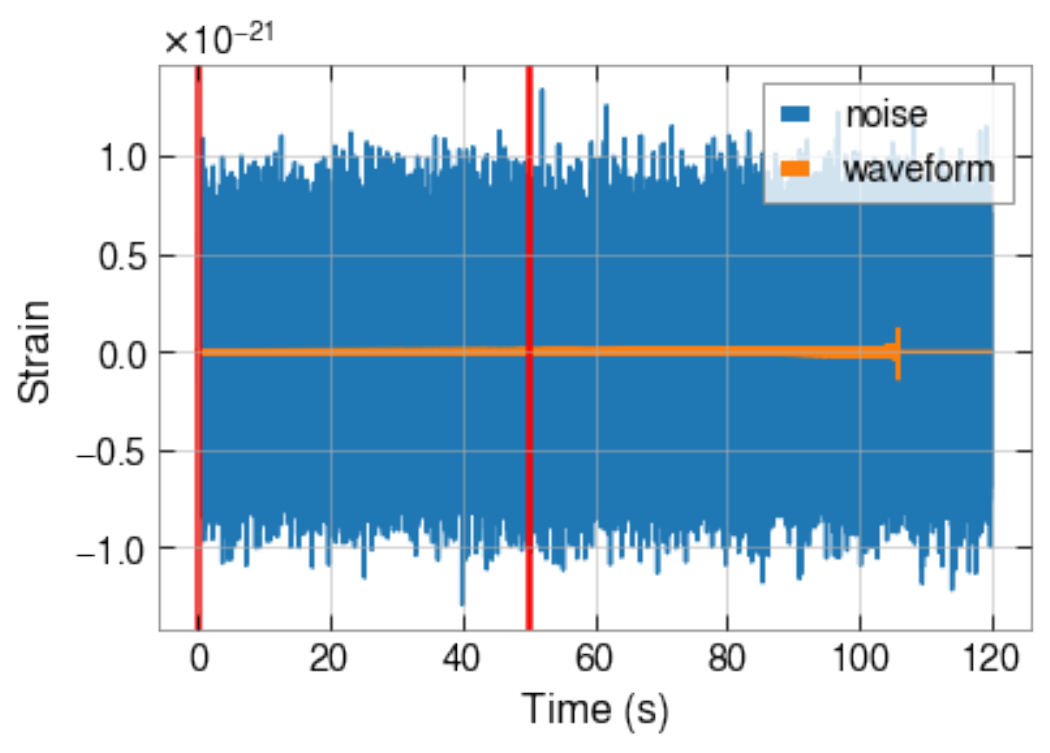}
    \caption{\label{fig:SigAndNoise} Representation of the noise and the injected waveform before the whitening. The CBC signal corresponds to a BNS where both component masses are $1.8\,M_{\odot}$ and the binary is placed at a luminosity distance of $100$ Mpc. At the time of training and testing of the CNNs, we do not pass this full frame to the network, but only the first 50 s (denoted by the two red lines), which is the chosen OTW length for this BNS category.}
\end{center}
\end{figure}

Since our objective is to train the networks on the early inspiral part of the waveforms, we select the desired OTW for the generated strain and its PI SNR is computed. In Fig.~\ref{fig:SigAndNoise}, we plot the waveform embedded in Gaussian noise. The vertical red lines represent the portion of the strain in the OTW. Note that the frame we use is always starts at the beginning of the 120 second injection it  is taken from. Finally, we whiten the stretch of data under consideration and normalise its amplitude by dividing all the points by the maximum amplitude in absolute value. Therefore, all points are in $[-1, 1]$. For the frames containing a GW, the event characteristics, such as the distance, are chosen so that the PI SNR distribution covers a wide range.

\section{Methodology}

\subsection{Architecture of the CNN}
The goal of this search is to perform a binary classification task, to distinguish the OTWs with GW signals from those without, with a short CNN, similarly to~\cite{George:2016hay, George:2017pmj, Gabbard:2017lja}. The CNNs were implemented with the \textit{PyTorch} package~\cite{NEURIPS2019_9015}. We use cross entropy as the loss function and  ADAMAX as the optimizer, which is a variant of ADAM, based on the infinity norm~\cite{kingma2019method}. Several hyperparameters such as the learning rate, the batch size, the numbers of layers, the kernel size, were  tested, but in this work we only report on the ones that provided the best performance.

After several trials, we found that the best performance with the minimum computational cost was acquired for $5$ convolutional layers. It was found that a bottleneck structure, i.e. starting with a large kernel size, making it smaller in the middle and enlarging it again afterwards, yielded the best results. We represent the best-performing architecture in Fig.~\ref{fig:config} and in Appendix A. The output of the network is a probability vector which contains the probabilities of the template belonging to the event class, where the event is present into the noise,  or to the noise class otherwise. The classification task is performed according to a predefined threshold, which is associated with the False Alarm Probability (FAP).


\begin{figure}[ht]
\begin{center}
    \includegraphics[width=0.55\textwidth]{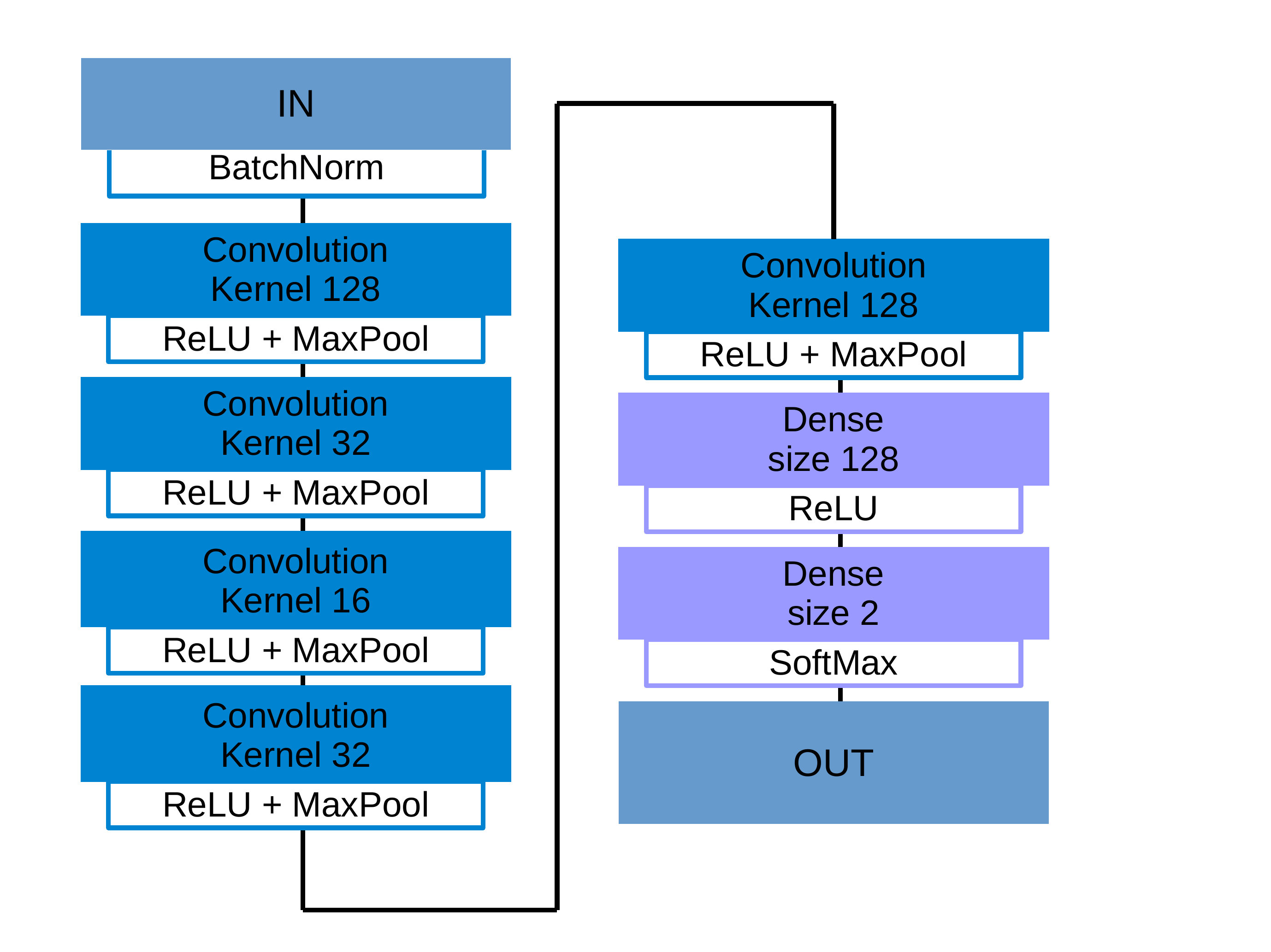}
    \caption{\label{fig:config} Architecture of the best performing CNN for all the categories. From one BNS type to the other, one needs to adapt the input size.}
\end{center}
\end{figure}

\subsection{Training and testing of our neural networks}


For each category we have a predefined OTW, given in Table~\ref{tab:inputs}. Due to the varying size of the inputs we perform a binary classification task with a tuned replication of the CNN for each BNS category. The data set is balanced, i.e. it contains 4000 frames of the noise + signal class, and 4000 frames of the noise class, where each of the frames corresponds to an OTW, built as described in the previous section. We employ $80\%$ of the data set for training and $20\%$ for validation. The performance of the network in the training and the validation sets is compared to avoid overfitting. Finally, we test the network with 2000 frames of the noise + signal class, where the events are chosen to fall into the distance range considered for each BNS category, and 2000 frames of the noise class. More information about the distributions of the data set can be found in Appendix B.

To assess the performance of each neural network, we classify its output for a given data frame into  true positives (TP), true negatives (TN), false positives (FP), and false negatives (FN), according to the standard \textit{confusion matrix}~\cite{Fawcett:2006}. 

We also define the True Alarm Probability (TAP) and the False Alarm Probability (FAP) as follows:
\begin{equation}\label{eq:TAP}
TAP = \frac{TP}{TP+FN} \qquad FAP = \frac{FP}{TN + FP}.
\end{equation}
The TAP corresponds to the number of noise + signal classified as such over all the number of frames that belong to the noise + signal class, whereas the FAP represents the number of noise frames which are misclassified over the number of frames that belong to the noise class. The performance of the networks will be evaluated based on the TAP for a fixed FAP, which is related to the threshold discussed in the previous subsection.

For this paper, we decided to present all the results for an FAP of $1\%$. This can be considered to be high if compared with the current GW searches, but we want to insist on the fact that this work is a proof of concept and that our pipeline uses only one detector. We expect that, by considering coincident triggers in $N_d$ detectors, the FAP will roughly go as $0.01^{N_{d}}$. This is an approximation where we assume that the three channels are independent and that, at each instant, each CNN has $1\%$ chance to claim a false detection.

\section{Results and discussion}\label{sec:Discussion}
In this section, we first discuss the performance of the three networks. Then, we report on the results of our method when applied to a realistic population of BNSs. Finally, we discuss a first attempt at curriculum learning, which is promising for the future.

\subsection{Performance of the CNNs}

\begin{figure}[t]
\begin{center}
    \includegraphics[keepaspectratio, width=0.5\textwidth]{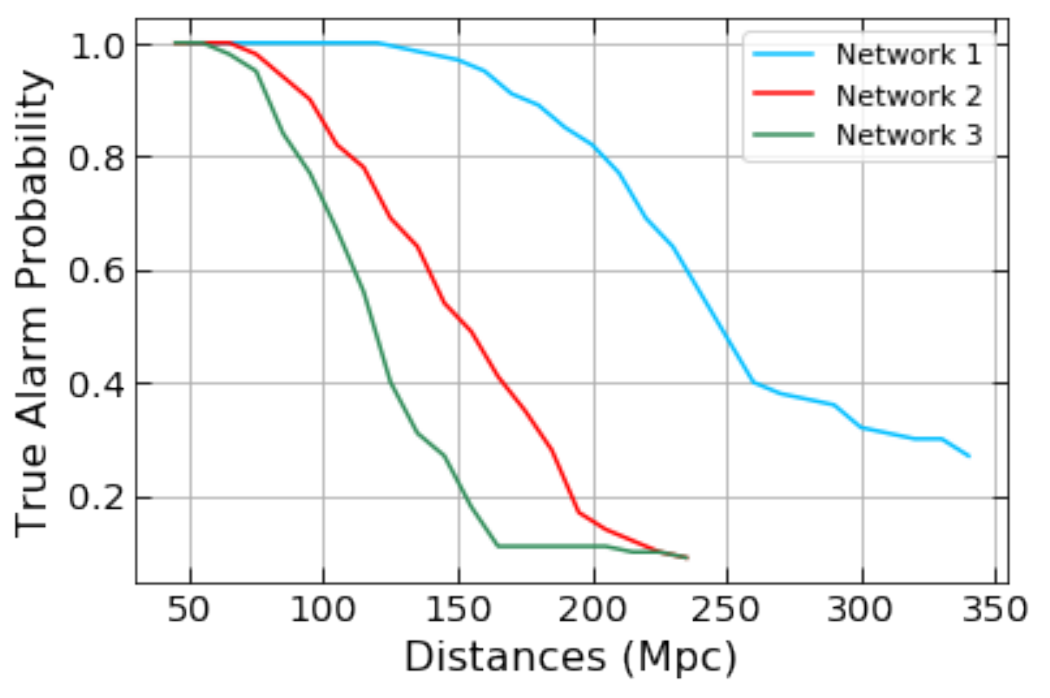}
    \includegraphics[keepaspectratio, width=0.5\textwidth]{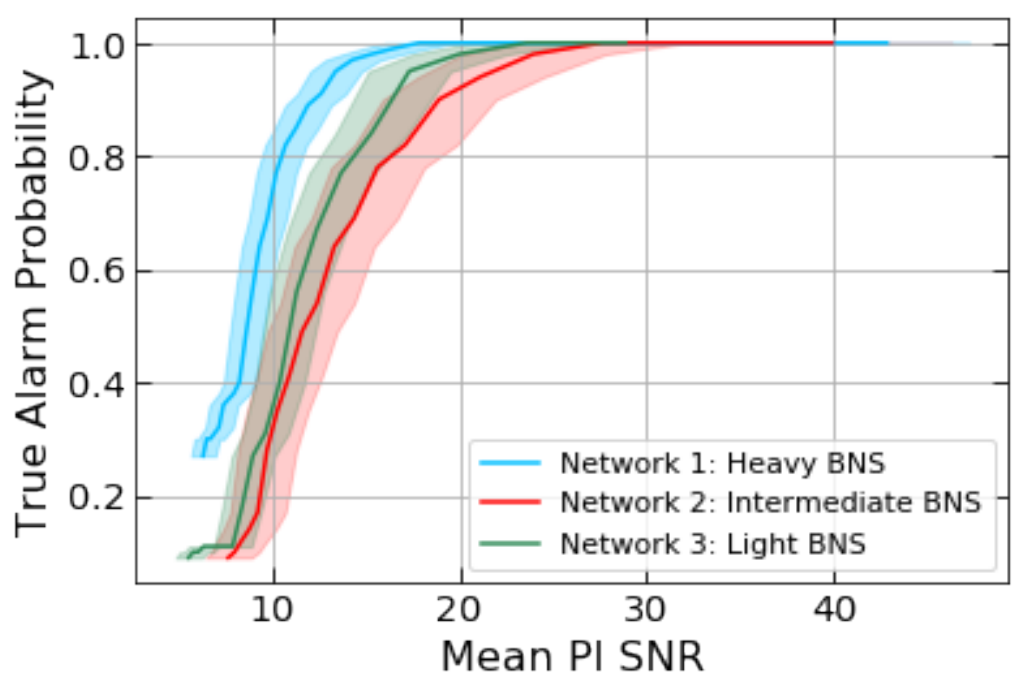}
    \caption{\label{fig:SumResults} 
   {The top panel represents the results of the three networks, each trained on its category, as a function of the distance. In the second panel, we compute the mean $\mu$ PI SNR and its standard error $\epsilon$, as $\mu(PI \text{ }SNR) \pm \epsilon(PI \text{ } SNR)$ for each distance and a confidence of $2 \sigma$, represented by the coloured band. For each graph the FAP is fixed at 0.01.}
    }
\end{center}
\end{figure}

In Fig.~\ref{fig:SumResults} we plot the TAP as a function of the distance in Mpc and the PI SNR for each category individually. We see that the network trained on heavier objects is able to reach higher distances. From Eq.~(\ref{eq:rho}) we can observe the same behaviour, as for smaller chirp masses we need to decrease the luminosity distance in order to keep the same SNR value. We obtain the best performance for the heavy BNS category. The intermediate and low categories have very similar performance, where we see that the $2\sigma$ interval of the two categories overlaps when considering the PI SNR. We also note that, since the architecture of the network has been optimised for the heavy category, it is expected that it performs best for this BNS category.

Note that the CNNs are sensitive to the accumulation of the signal. To confirm this, we trained and tested the networks on data with low frequency cut-offs set to higher values than the usual $20$~Hz. This is a way to reduce the PI SNR of the injected signal while maintaining the same maximum amplitude. For the testing set, we obtained an $88\%$ TAP for a cut-off at $20\,$Hz, and $71\%$ for a low-pass frequency at $26\,$Hz, showing that the CNN is sensitive to the PI SNR for a fixed maximum amplitude. Similarly to matched filtering, a CNN is designed to recognise patterns and, in this context, the larger PI SNR means that the signal is present for a longer time. 

From Fig.~\ref{fig:SumResults}, we see that Network 1 is able to reach distances larger than 60~Mpc before its TAP has a departure from 100\%. As the first BNS detected GW170817 was located at a distance of the order of 40~Mpc, our method will have a high probability to detect similar signals when present in noise at design sensitivity. 
This means that our method is able to recover realistic signal from Gaussian noise when only the inspiral part is present.
Network 2, which is trained on intermediate BNSs, is able to have a better performance at higher distances, which is expected based on the chirp mass - PI SNR relation. Finally, Network 3 has a TAP of $100\%$ even for a distance of 125~Mpc, meaning that the efficiency of detection is still high for distances similar to that of GW190425, the second BNS discovered by the LIGO-Virgo collaboration~\cite{Abbott:2020uma}. 

\begin{figure}[ht]
\begin{center}
    \includegraphics[keepaspectratio, width=0.42\textwidth]{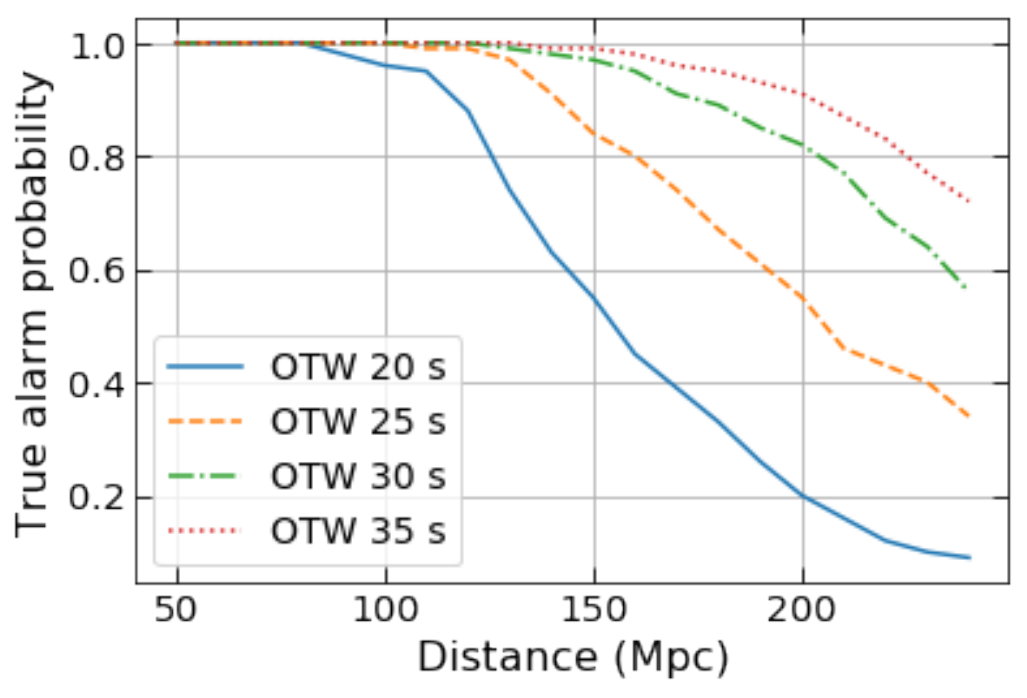}
    \caption{\label{fig:TW_evol} Representation of the performance of the CNN trained on the heavy BNS systems for different OTWs. One sees that a longer window gives a higher number of detections. However, it also means the detection happens closer to the merger time. The mean times before merger are 35, 30, 25, and 20 seconds for the 20, 25, 30, and 35 seconds OTW, respectively.}
\end{center}
\end{figure}

We now perform a series of tests to evaluate the influence of the length of the OTW. Indeed, this is an important hyperparameter that represents the fraction of the signal seen for a given event. It needs to be optimised to have as many detections as possible while keeping a long enough delay between the trigger and the merger time. In Fig.~\ref{fig:TW_evol}, we show the TAP for Network 3 when using different OTW. As expected, a larger OTW  increases the TAP, but is associated with a shorter time lapse before the merger.

We also test whether a network trained on a certain category is able to find signals that belong to a different category. We concentrate on Network 3, which is trained to detect heavy BNSs, and check whether it is capable of detecting intermediate BNSs. For this, we increase the OTW of intermediate BNSs to 30 s, to be able to feed the data set to Network 3. We find that the TAP decreases significantly. Network 2, which is trained to detect intermediate BNSs, yields a TAP of $\sim 68\%$, while Network 3 reaches only $\sim 16\%$.  This is also understandable in terms of PI SNR, as the reduction of the OTW duration leads to a decrease in the PI SNR, and we already established that this is the key parameter for detection.


We now compare the time needed for our CNN to analyse one frame with the time needed for matched filtering. When applying matched filtering on a 50 s frame, similar to those passed to the CNNs and with only the optimal template, the computation time is $\sim 0.05~s$.  This is just the bare minimum time needed to get the SNR in matched filtering. In this traditional method, several templates are tested and the trigger is not only assigned an SNR, but also other statistics, such as the FAR. As a consequence, the time to get the final information is longer~\cite{sachdev2019gstlal}. Analysing the same frame using our CNN on a \textit{Nividia GeForce RTX 2070 SUPER} GPU, we get the probability of an inspiral to be present in $\sim 0.005\, s$. Therefore, the time needed to analyse the frame and get a prediction probability is improved by a factor of $10$. 


\subsection{Test on a realistic population of BNS}

In order to have a better grasp on the performance of our networks with respect to matched filtering, we also test them on a simulated realistic population of BNS systems. Therefore, we compute both the optimal SNR and the PI SNR for each BNS with a high-frequency cut-off of 32 Hz, similarly to what was done in ~\cite{Sachdev:2020lfd}. When performing the run with a high-frequency cut-off and the test with our CNNs, we only consider the events that have a full matched filtering SNR higher than $8$. This basic computation is performed for the high rate presented in ~\cite{Abbott:2020gyp} 

The cut-off frequency of 32 Hz has been chosen to give results comparable to those in~\cite{Sachdev:2020lfd}, while having in-band times that correspond to the OTWs defined in Table~\ref{tab:inputs}.

\begin{figure}
    \centering
    \includegraphics[keepaspectratio, width=0.5\textwidth]{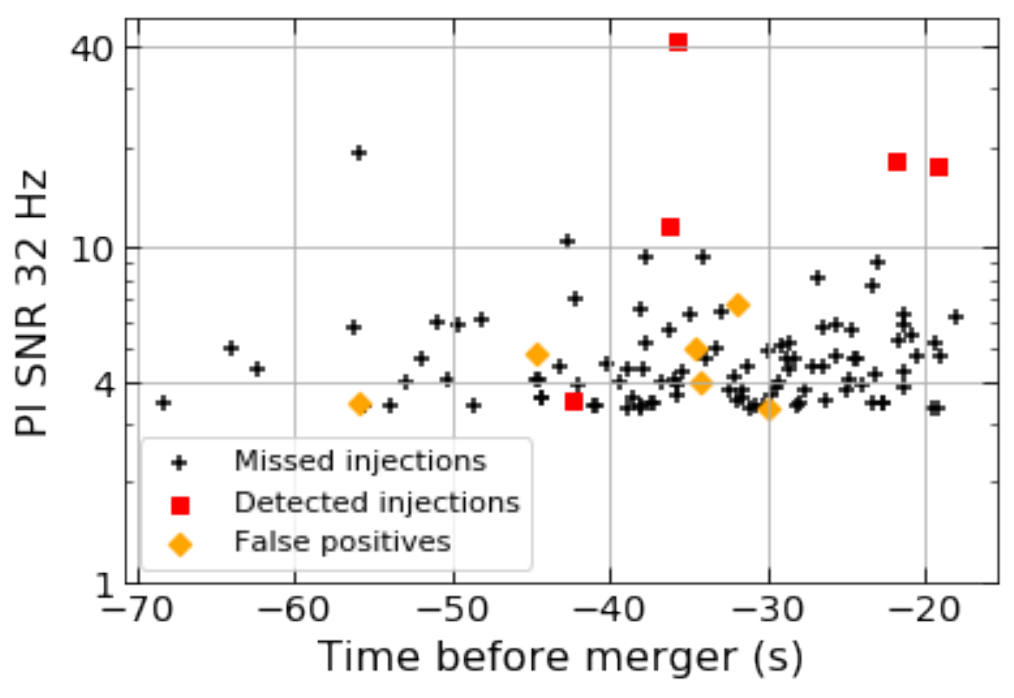}
    \caption{The PI SNR for a low-pass filter at 32~Hz for each BNS with a full SNR higher than 8. The black crosses represent the events missed by the CNNs, and the red squares are the events correctly found. The orange diamonds are triggers that correspond to noise fluctuations (false positives).}
    \label{fig:PropDetectBNSpop}
\end{figure}

The population synthesis is performed using the code of ~\cite{Samajdar:2021egv}, with minor changes in order to suit our framework. For example, the PSD employed is the same as for the noise generation, the low frequency cut-off used is 20~Hz, and we generate the equivalent of 5 years of data. 

One shortcoming of this procedure to generate a realistic population of BNSs is that, although it is fast, it is based on analytical approximations. As a consequence, we do not inject the signals in noise to compute the SNR, and are not able to compute the matched filtering false alarm rate (FAR) for such frames. So, we cannot use the criterion of Ref.~\cite{Sachdev:2020lfd} (namely an SNR threshold followed by an FAR) and the direct comparison is non-trivial. Our procedure confirms the difficulty to detect those events with matched filtering methods.

Once we have selected the events based on the analytical approach, we inject them in design-sensitivity noise and pass the frames to the CNNs. Fig.~\ref{fig:PropDetectBNSpop} represents the events detected, those missed and those misidentified. We also generate the same noise for each event but without injecting the BNS in it. We test our networks on these pure noise frames to highlight the false positives. As shown in Fig.~\ref{fig:SumResults}, the networks detect most of the BNSs which have a PI SNR sufficiently high. 
We also want to emphasise the fact that matched filtering applied for pre-merger alert also needs the PI SNR to be above a threshold to lead to a trigger. This threshold depends of the framework, and the number of detectors included. We can see that, if one chooses an SNR threshold of 8, our results are comparable to those of matched filtering. Nevertheless CNNs are much faster\footnote{Here, we neglected the latency needed for the data transfer. It would be, in the worst case, comparable to that of~\cite{Sachdev:2020lfd}.}.

A key feature employed in ~\cite{Sachdev:2020lfd} is the network of detectors. Requiring coincident detections in the different detectors helps to remove signals due to noise artefacts. Another advantage is that the signal can accumulate in several detectors simultaneously. Additionally, the sky localisation is found using the data in the three detectors~\cite{Aasi:2013wya}.  For a neural network, the input will have a certain number of channels, one for each detector. Then, the input will be convolved through the network, finding relationships between the different channels. This should decrease the FAP of our detector network and enable us to find the sky localisation. This will be explored in a future work. 

\subsection{Basic curriculum learning exploration}

Aside from the architecture, another key factor in the development of DL algorithms is the training procedure. From the population analysis we conclude that the networks see the loudest events, i.e. those with the highest PI SNR in the OTW (or the highest SNR in the detector for the full template). The networks have been trained on a very wide distance range for the events (hence a wide PI SNR range), but it is hard for them to detect smaller PI SNR, as we can see in Fig.~\ref{fig:SumResults}. A way to overcome this obstacle is by training the CNNs with \textit{curriculum learning}. The main idea is to train the network on batches of PI SNR, first on the easy examples, namely the frames with highest PI SNR. Then the difficulty is increased iteratively by decreasing the PI SNR, until the hardest examples are reached, namely the frames with the lowest PI SNR (see~\cite{George:2017pmj} or~\cite{portilla2020deep} for an example).

\begin{figure}
    \centering
    \includegraphics[keepaspectratio, width=0.5\textwidth]{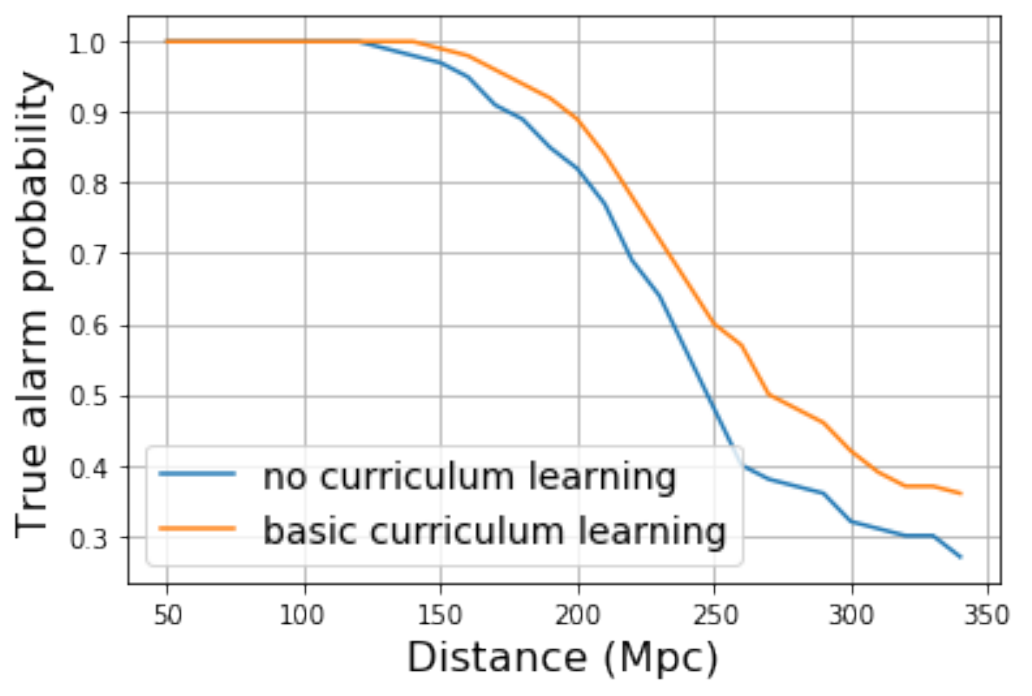}
    \caption{Comparison of the TAP as a function of the distance for the GW sources with and without basic curriculum learning for the heavy BNS class. One sees that even a very rudimentary curriculum learning setup helps improve the TAP at higher distances. Note that the blue curve is the same than in Fig.~\ref{fig:SumResults}}
    \label{fig:CLtest}
\end{figure}

With this idea in mind, we generate an extra batch of training data with higher distances and lower PI SNR. Thus, we train on the first data set, store the weights, then train on the newly generated set starting with the previously stored weights. The results of this test can be seen in Fig.~\ref{fig:CLtest}. It can be observed that the TAP increases significantly even if we are using only one extra data batch. As a consequence, we expect the efficiency of our networks to increase substantially once they are trained through the curriculum learning methodology.


\section{Conclusion}\label{sec:Conclusions}

In this work, we have introduced a new approach based on short CNNs for pre-merger alert. We have shown that it is possible to detect BNS events when only part of the early inspiral is present in the data stretch under consideration. For this purpose, we have introduced three different neural networks, each trained on a particular range of chirp masses for the BNS systems. Such developments are important in the context of MMA, as the prediction stage is computationally less expensive and usually faster than for traditional matched filtering. We have also shown that our method is able to recover signals coming from a realistic BNS population simulated at design sensitivity, and compared our detection statistics to those obtained with current matched filtering pipelines. In addition, we also suggested some improvements in the training method, as well as in the structure of our CNNs, to enhance their performance further, leading the way to a pre-merger alert system that would be competitive.

This paper was presented as a a proof of concept and we will continue to build upon this basis to upgrade our networks and get an even better performance. 
The next steps, which will probably require more complex networks, are the consideration of multiple interferometers and the implementation of sky localisation. Furthermore, curriculum training will be systematically deployed, as this will allow us to train on a bigger dataset with smaller PI SNR. Indeed, the training set currently has a minimum PI SNR around 8. With curriculum learning it will be possible to lower this value. 
A fourth CNN trained to retrieve the full BNS signal regardless of the category will be built. This will complete the pipeline as the events that are not detected based only on their inspiral would still be found in low-latency. 

It has already been shown in various works that ML-based algorithms can help GW astronomy. In this work, we have shown that it can also be used to solve one of the challenges that will arise in the future, namely the early detection of BNS mergers in the context of MMA. However, we still want to improve the performance and add some features, such as sky location. These are the next milestones which will probably require more complex networks and more advanced training methods.\\

\section*{Acknowledgments}
The authors thank Michal Bejger, Chris Messenger and Andrew Miller for their useful comments, as well as Maxime Fays, Vincent Boudart and Chris Van Den Broeck for useful discussions. G.B. is supported by a FRIA grant from the Fonds de la Recherche Scientifique-FNRS, Belgium. J.R.C. acknowledges the support of the Fonds de la Recherche Scientifique-FNRS, Belgium, under grant No. 4.4501.19. M.L., S.C, A.R, and J.J  are supported by the research program of the Netherlands Organisation for Scientific Research (NWO). The authors are grateful for computational resources provided by the LIGO Laboratory and supported by the National Science Foundation Grants No. PHY-0757058 and No. PHY-0823459.


\bibliographystyle{apsrev}
\bibliography{Manuscript}{}

\appendix

\onecolumngrid


\section{Details on the architecture of the CNNs}

\begin{figure}[h!]
    \centering
    \includegraphics[keepaspectratio, width=0.5\textwidth]{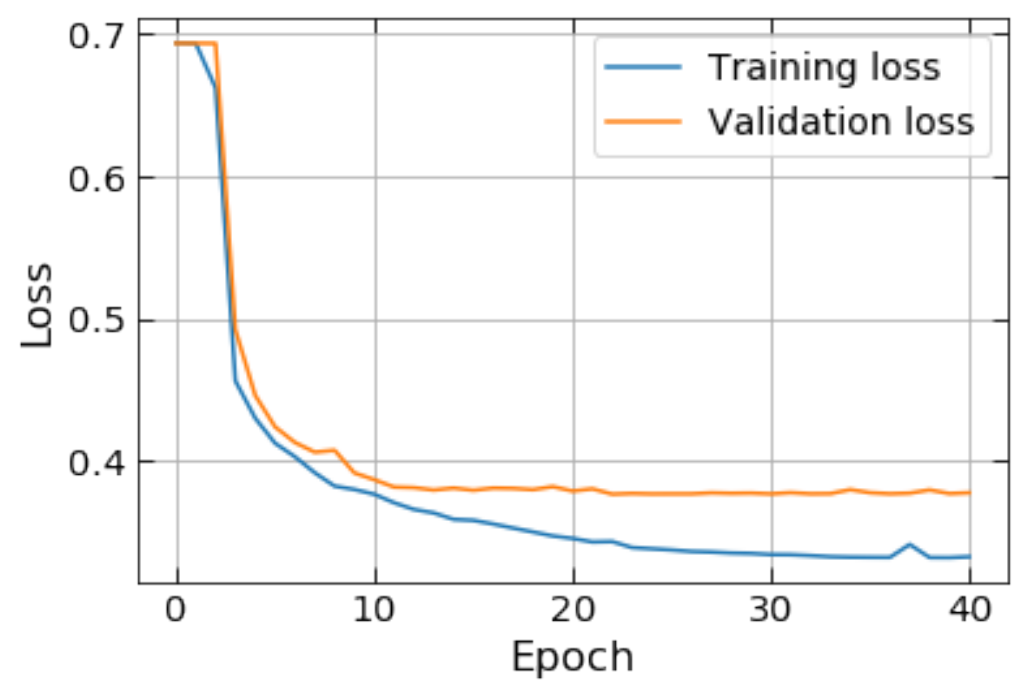}
    \caption{The loss variation for the training set and the validation set of CNN 3 as a function of the epoch.}
    \label{fig:loss}
\end{figure}

In Section III we briefly described the networks, but in this section we provide with more details about the architecture (see Table ~\ref{tab:DetailsLayers}), and the different hyper-parameters fine-tuned for  heavy BNS category. 
 The batch size of the training was 40 for networks 1 and 2, and 30 for network 3, due to memory issues. We employ the cross-entropy as loss-function. The optimizer is Adamax \cite{kingma2019method} with a learning rate of $8 \times 10^{-5}$ and a weight decay of $10^{-5}$. We trained the networks over 40 epochs. Usually, the validation and training loss drop before epoch 5, as we show in  Fig.~\ref{fig:loss}, where plot the training and validation loss of  the neural network 3, on  heavy BNS category. As a consequence, and to avoid over-fitting, we generally use early stopping (around the $12^{th}$ epoch).

The training of the three networks was done with a dataset of 8000 frames. The data sets are balanced so that half of them correspond to noise and the other half are noise + waveform. The testing was performed with a testing set of 4000 frames, where again, half of them correspond to noise, the other half are noise + waveform.

\begin{table*}[htbp]
\centering
\begin{tabular}{|c|c|c|c|c|c|c|c|}
  \hline
   Layers & Input & Output & Kernel size & Stride & Padding & Dilation & Activation  \\
  \hline
  \hline
   BatchNorm & 1  & 1 & - & - & - & - & - \\\hline
   
   Conv1D & 1 & 32 & 128 & 1 & 0 & 1 & ReLU \\\hline
   MaxPool1D & 32 & 32 & 4 & 4 & 0 & 1 & - \\\hline
   
   Conv1D & 32 & 64 & 32 & 1 & 0 & 1 & ReLU \\\hline
   MaxPool1D & 64 & 64 & 4 & 4 & 0 & 1 & - \\\hline
   
   Conv1D & 64 & 128 & 16 & 1 & 0 & 1 & ReLU \\\hline
   MaxPool1D & 128 & 128 & 4 & 4 & 0 & 1 & - \\\hline
   
   Conv1D & 128 & 256 & 32 & 1 & 0 & 1 & ReLU \\\hline
   MaxPool1D & 256 & 256 & 4 & 4 & 0 & 1 & - \\\hline
   
   Conv1D & 256 & 612 & 128 & 1 & 0 & 1 & ReLU \\\hline
   MaxPool1D & 612 & 612 & 4 & 4 & 0 & 1 & - \\\hline
   
   Dense & $X$ & 128 & -  & - & - & - & ReLU \\\hline
   Dense & 128 & 2 & -  & - & - & - & SoftMax \\\hline
\end{tabular}
\caption{Complete architecture of our CNNs. Between the last MaxPool1D layers we flatten all the channels to obtain an output of dimension 1 and length $X$ (the $X$ depends of the OTW).}
\label{tab:DetailsLayers}
\end{table*}

\section{Data distribution}

\begin{figure*}[htbp]
  \begin{subfigure}[b]{0.5\linewidth}
    \centering
    \includegraphics[width=0.9\linewidth]{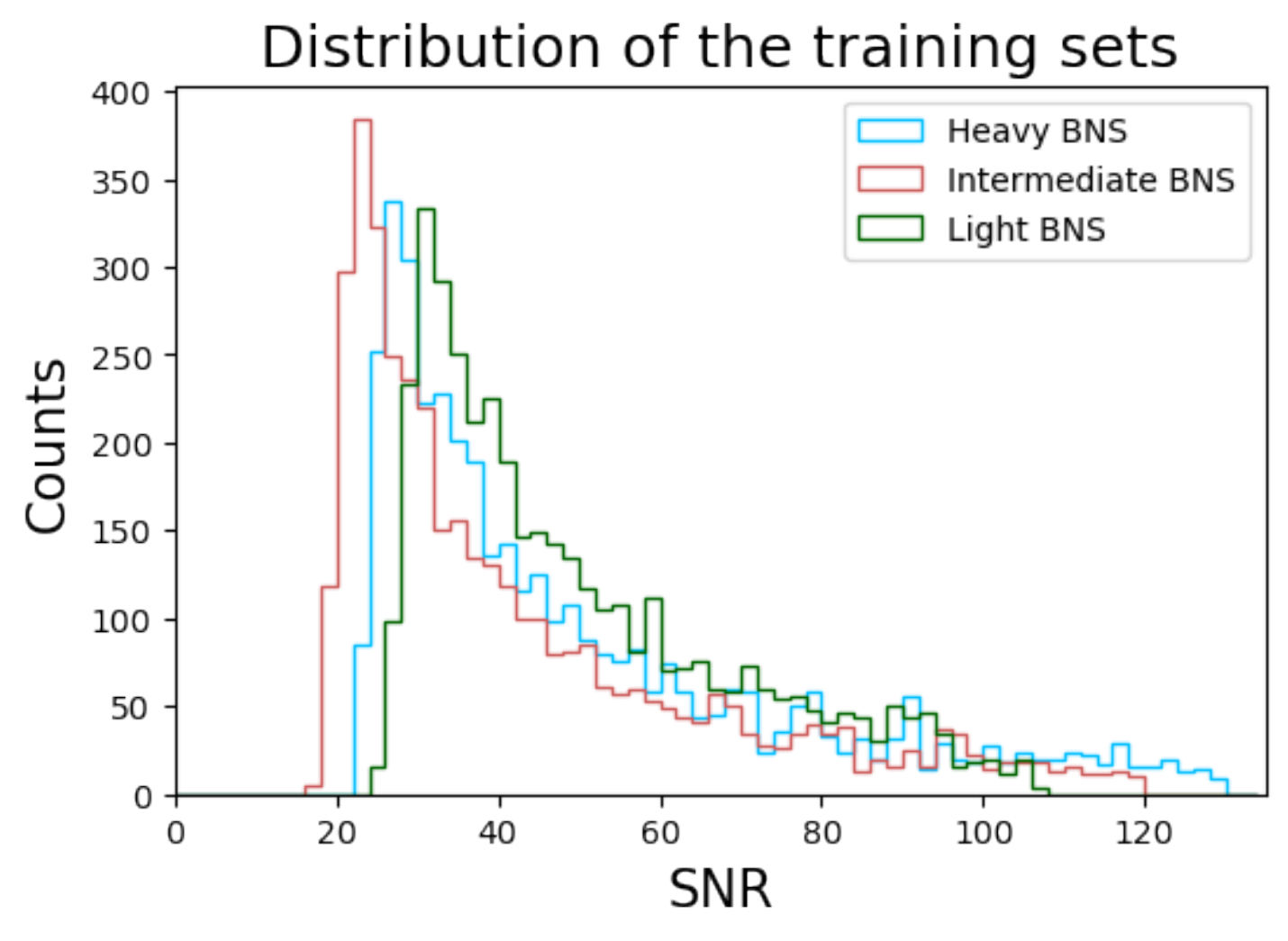} 
    \caption{Distribution of SNR for the training sets.} 
    \label{fig7:a} 
    \vspace{4ex}
  \end{subfigure}
  \begin{subfigure}[b]{0.5\linewidth}
    \centering
    \includegraphics[width=0.9\linewidth]{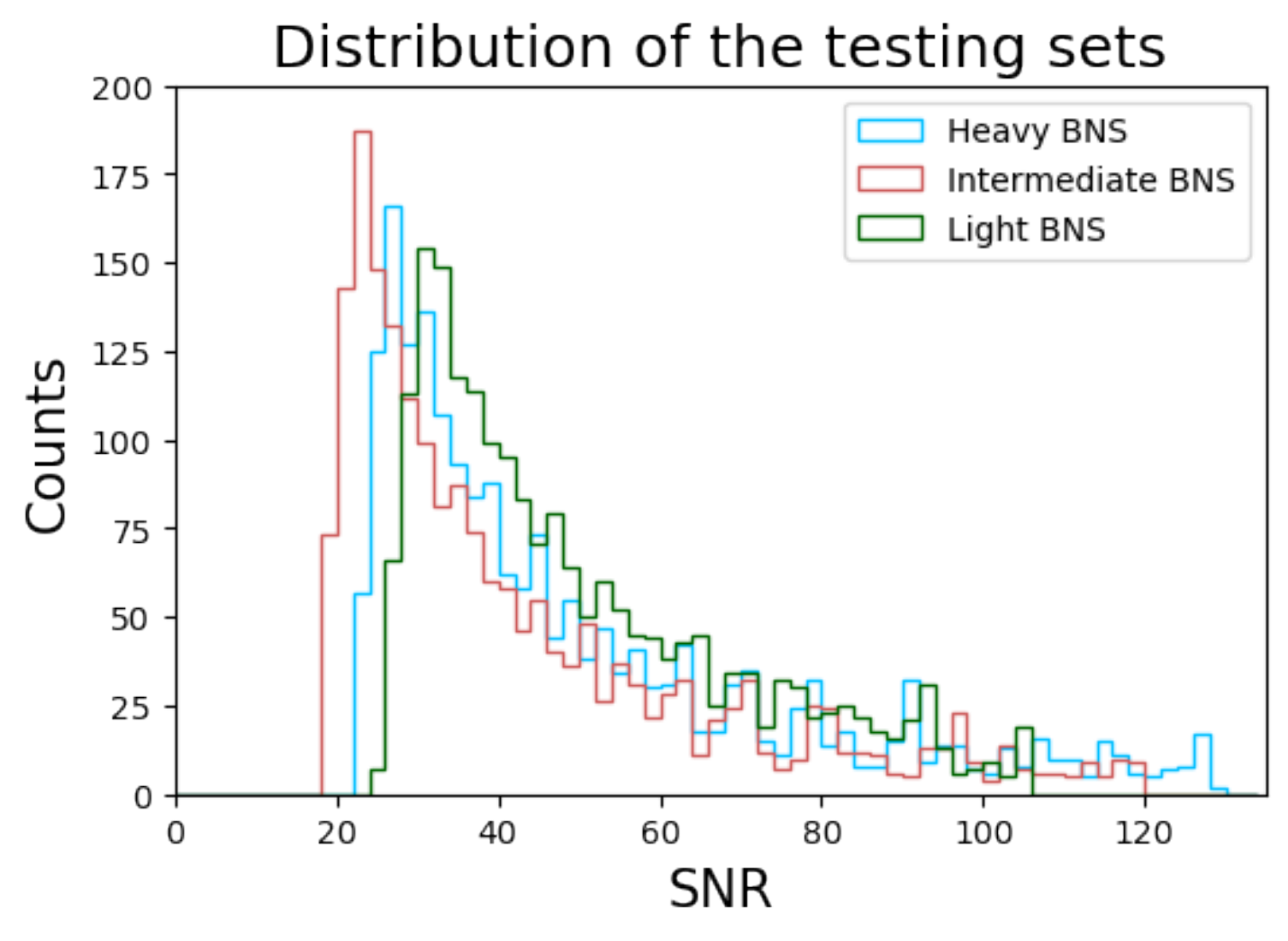} 
    \caption{Distribution of SNR fot the testing sets.} 
    \label{fig7:b} 
    \vspace{4ex}
  \end{subfigure} 
  \begin{subfigure}[b]{0.5\linewidth}
    \centering
    \includegraphics[width=0.9\linewidth]{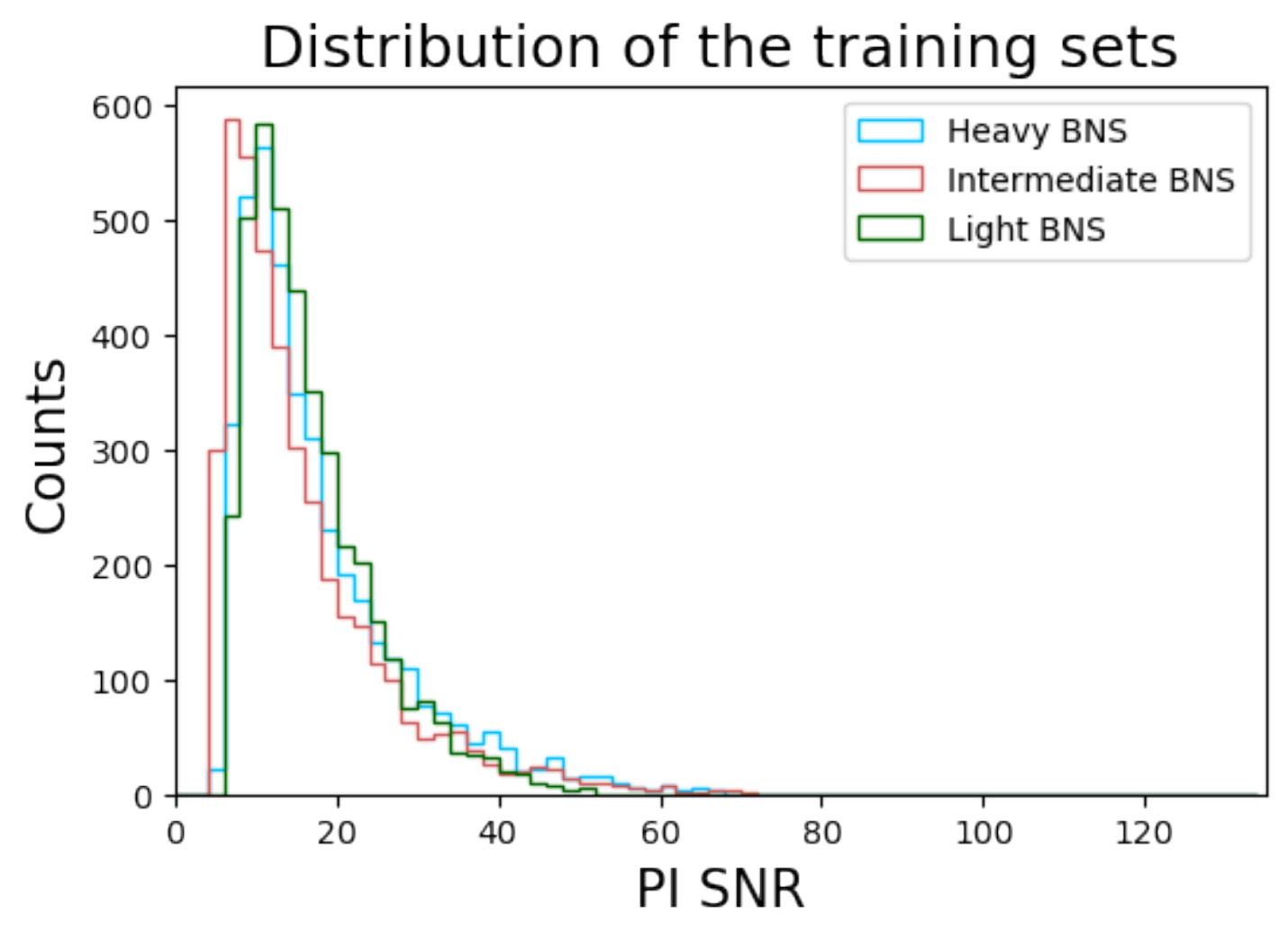} 
    \caption{Distribution of PI SNR for the training sets.} 
    \label{fig7:c} 
  \end{subfigure}
  \begin{subfigure}[b]{0.5\linewidth}
    \centering
    \includegraphics[width=0.9\linewidth]{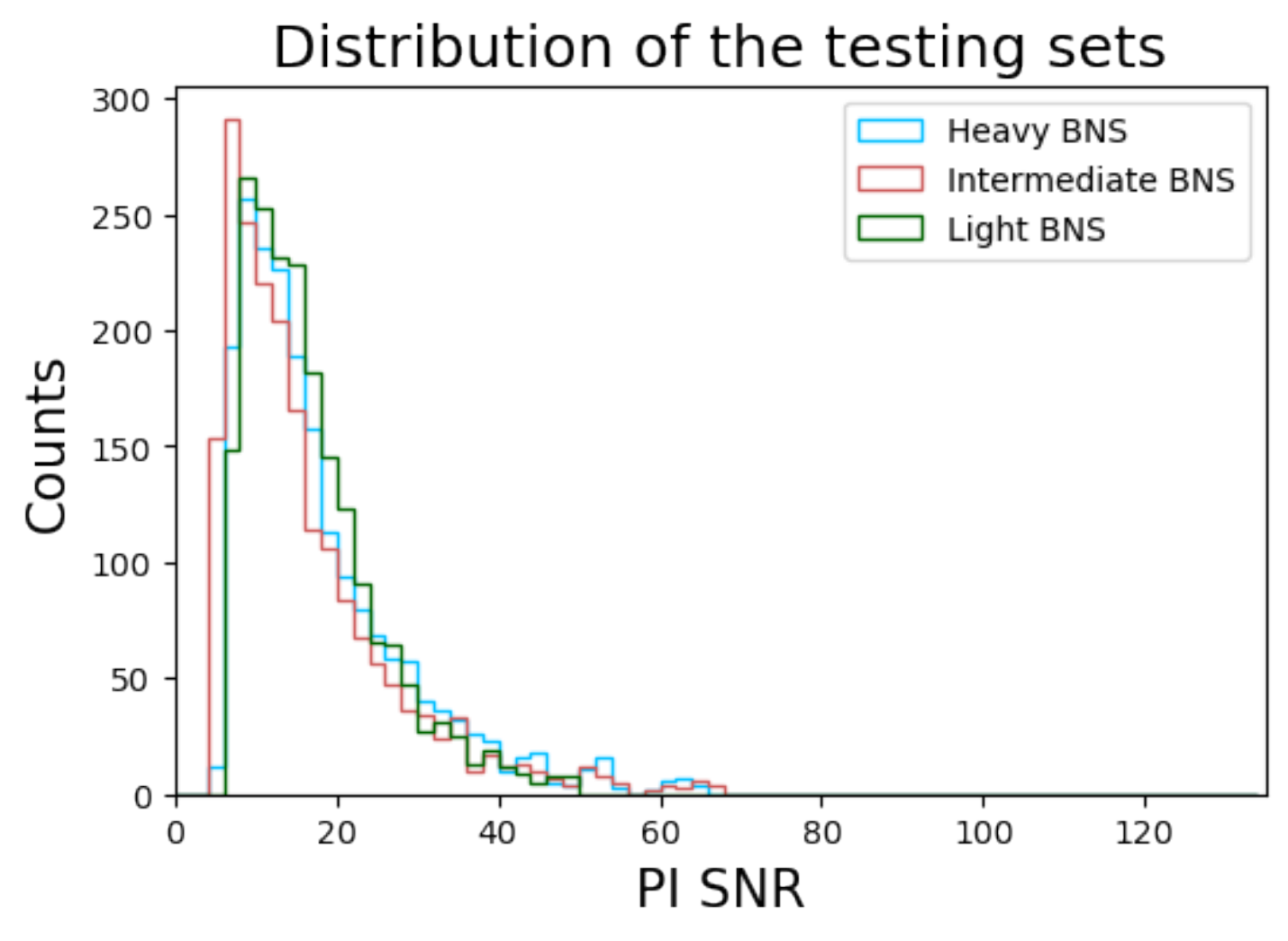} 
    \caption{Distribution of PI SNR for the testing sets.} 
    \label{fig7:d} 
  \end{subfigure} 
  \caption{Data distributions as functions of the SNR and PI SNR.}
  \label{fig:distributions} 
\end{figure*}

In this section we represent the data distribution with respect to the to SNR and PI SNR. Each distribution employed for training contains 4000 frames, and each distribution employed for testing contains 2000 frames. In Fig.~\ref{fig:distributions} we observe that the main difference between the data distributions against SNR or PI SNR is a shift and a decrease in the range of PI SNR due to the removal of the merger from the frames. Indeed, the SNR is in the range $\approx [20, 130]$, while the PI SNR is in the range $\approx [1, 70]$. Therefore, due to the smallness of the PI SNR, the classification task becomes more difficult.

\twocolumngrid

\end{document}